\documentclass{aa}
\usepackage{graphicx} 
\usepackage{orcidlink}
\usepackage{amsmath}

\usepackage{pgf}

\usepackage{txfonts}

\usepackage{units}
\usepackage{pdflscape}
\usepackage{makecell}

\usepackage{hyperref} 
\hypersetup{
  colorlinks=true,
  linkcolor=blue,
  citecolor=blue,
  urlcolor=blue
}
\begin{document}
\title{CHANG-ES}
\subtitle{XXXVII. Effects of spectral aging on radio scale heights}
\author{D.~C.~Smolinski\inst{1}
\and
V.~Heesen\inst{1}
\orcidlink{0000-0002-2082-407X}
\and 
M.~Br\"uggen\inst{1}
\orcidlink{0000-0002-3369-7735}
\and
J.-T.~Li\inst{2}\orcidlink{0000-0001-6239-3821}
\and
M.~We\.zgowiec\inst{3}\orcidlink{0000-0002-4112-9607}
\and 
M.~Stein\inst{4}\orcidlink{0000-0001-8428-7085}
\and
L.-Y. Lu \inst{5}\orcidlink{0000-0002-3286-5346}
\and
T. Wiegert\inst{6}\orcidlink{0000-0002-3502-4833}
\and
J. Irwin\inst{7}\orcidlink{0000-0003-0073-0903}
\and
R.-J. Dettmar\inst{4}\orcidlink{0000-0001-8206-5956}
}
\titlerunning{Scale heights of the thick radio disc}
\authorrunning{D. C. Smolinski et al.}
\institute{Hamburger Sternwarte, Universität Hamburg, Gojenbergsweg 112, D-21029 Hamburg, Germany
            \email{david.smolinski@hs.uni-hamburg.de}
            \and
            Purple Mountain Observatory, Chinese Academy of Sciences, 10 Yuanhua Road, Nanjing 210023, China
            \and
            Obserwatorium Astronomiczne Uniwersytetu Jagiello\'nskiego, ul. Orla 171, 30-244 Kraków, Poland
            \and
            Ruhr University Bochum, Faculty of Physics and Astronomy, Astronomical Institute (AIRUB), 44780 Bochum, Germany
            \and
            Department of Physics and Astronomy \& Research Center of Astronomy, Qinghai University, 251 Ningda Road, Xining, 810016, China
            \and
            Instituto de Astrofısica de Andalucıa (IAA-CSIC), Glorieta de la Astronomıa, 18008, Granada, Spain
            \and
            Department of Physics, Engineering Physics \& Astronomy, Queen’s University, Kingston, ON K7L 3N6, Canada
            }

\date{Received October 2, 2025; accepted December 22, 2025}

\abstract
{Cosmic rays and magnetic fields play an important role for the formation and evolution of galaxies. Radio continuum observations allows us to study them in the haloes of edge-on galaxies.}
{We explore  
the frequency dependence of the radio scale height which depends on cosmic ray transport and electron cooling. We test the influence of fundamental galaxy properties, such as star-formation rate (SFR), mass and size.
}
{We used radio continuum data of 16 edge-on galaxies from the Continuum HAloes in Nearby
Galaxies – an EVLA Survey (CHANG-ES). We included maps from the LOw Frequency ARray at $\unit[144]{MHz}$ and from the Jansky Very Large Array at $\unit[3]{GHz}$ with $7\arcsec$ angular resolution. We extracted vertical intensity profiles within the effective radio radius and fitted beam-convolved double-exponential models to separate thin and thick discs. For the thick radio discs, we computed mean spectral indices and scale-height ratios between $\unit[144]{MHz}$ and $\unit[3]{GHz}$.}
{We find a mean scale-height ratio of $1.26\pm 0.16$. This is much lower than what we would expect for either cosmic ray diffusion or advection if synchrotron and inverse Compton losses dominate for the electrons. There is a moderate positive correlation between the ratio and spectral index of the thick disc: galaxies with high ratios have flat radio spectra. The ratio does not depend on any other galaxy parameter. The radio spectrum of the  thick disc, as indicated by the radio spectral index, steepens with total mass (strong correlation) and flattens with SFR-to-mass surface density (moderate correlation).}
{Galaxies with galactic winds have flat radio continuum spectra and large scale heights at low frequencies. This shows effective transport of cosmic rays in such systems.}


\titlerunning{The thin and thick radio disc in galaxies}
\authorrunning{D.~C.~Smolinski et al.}

\maketitle

\section{Introduction}

Understanding the evolution and formation of galaxies is highly complex and not yet fully understood.
Galaxies assemble through the accretion of gas during the early stages of cosmic history.
This gas subsequently condensed into stars that are found today in the bulge and the stellar halo. The discs in galaxies are typically a later addition, developing as material with higher angular momentum drifts into the galaxies \citep{ostriker_theoretical_2012}. This has been the prevailing theoretical framework for galaxy formation. However, recent James Webb Space Telescope (JWST) observations have revealed surprisingly massive galaxies at high redshifts that potentially challenge aspects of current galaxy formation models \citep{boylan-kolchin_stress_2023}.\\
These findings further highlight the importance of understanding the complex baryonic processes that regulate galaxy growth and evolution. In this context, the circumgalactic medium (CGM) plays a crucial role since it forms the interface between the galaxies and their large-scale environment. Multiple processes within the CGM shape galaxy evolution.  Accretion continuously supplies gas from the intergalactic medium and the infall of  satellite galaxies contribute further material to the system \citep{hafen_origins_2019}. Thermal instabilities facilitate transitions between gas phases. Galactic winds redistribute disc gas throughout the halo. Understanding these individual processes and their interplay in the CGM is crucial for developing a comprehensive picture of galaxy evolution. The CGM directly fuels star formation while simultaneously moderating feedback processes \citep{tumlinson_circumgalactic_2017}. Detailed analysis of CGM properties – including its multi-phase structure, gas dynamics, and metallicity – requires examining, both, thermal and dynamical processes as well as non-thermal components such as magnetic fields and cosmic rays \citep[CRs;][]{ji_properties_2020, van_de_voort_effect_2021}. It has been shown that CR affect the accretion of gas on galaxies by modifying the CGM flow structure. Therefore CGM densities and temperatures can be strongly affected by CR \citep{buck_effects_2020}. Magnetic fields act as a pressure component in the CGM which needs to be considered when determining the dynamics and structure of the CGM. Thus, magnetic fields change the properties of the CGM significantly \citep{pakmor_magnetizing_2020, van_de_voort_effect_2021}.\\
In order to probe the CGM, radio continuum observations have proven particularly valuable. With polarised background radio galaxies, magnetic fields were detected in the CGM of highly inclined galaxies using Faraday rotation \citep{heesen_nearby_2023}. Such edge-on galaxies reveal in the radio continuum emission two distinct discs. The thin disc, with scale heights of a few hundred parsecs, consists mostly of stars, dust, and gas \citep{ferriere_interstellar_2001}. The thick disc, on the other hand, comprises mostly ionized gas, cosmic rays, and magnetic fields \citep{cox_three-phase_2005}. The thick disc is inflated by non-thermal pressures and has a different kinematic and chemical composition than the thin disc. The thick gaseous disc is the location where galactic in- and outflows can be studied. Therefore, the thick disc serves as an interface between the galaxy and the CGM. Thus, the thick disc can be used to constrain galaxy formation models \citep{tsukui_emergence_2025}.\\
Given the thick disc's importance as an interface, radio continuum observations represent one of the most essential tools for understanding the influence of CRs and magnetic fields on galaxy evolution since the radio continuum is mainly made up by synchrotron radiation. 
However, disentangling the individual contributions of these components remains challenging, as synchrotron emission depends on, both, the cosmic ray population and the magnetic field strength. Radio continuum observations are particularly valuable for studying CRs, which play a dynamical role in galaxy evolution. CRs can drive galactic outflows by exerting vertical pressure gradients without significantly heating the gas \citep{heintz_role_2020}. Although only $\sim$10\% of a supernova remnant's kinetic energy is converted into 
acceleration of CRs, they have a disproportionate impact on galactic dynamics. This is because CRs are transported away from their injection sites in star-forming regions and subsequently establish vertical pressure gradients that drive galactic winds \citep{girichidis_cooler_2018}.\\

With the Continuum HAlos in Nearby Galaxies – an EVLA Survey (CHANG-ES), it is possible to study the role of CRs in galaxy evolution using a sample of 35 edge-on galaxies. CHANG-ES is a radio continuum survey that observed these galaxies in $L$- ($\unit[1-2]{GHz})$, $S$- ($\unit[2-4]{GHz}$) and $C$-band ($\unit[4-6]{GHz})$, with full polarisation data available. The survey has already yielded important insights into the structure and dynamics of the extra-planar interstellar medium, cosmic ray transport, and magnetic field configurations in nearby galaxies.
In particular, Faraday rotation analyses have revealed large-scale magnetic field structures such as kiloparsec-scale magnetic ropes in NGC~4631 \citep{mora-partiarroyo_chang-es_2019} and a quadrupolar field geometry in NGC~4666 \citep{stein_chang-es_2019}, consistent with expectations from dynamo theory \citep{henriksen_galactic_2022}. A systematic Faraday rotation study across the entire CHANG-ES sample confirmed the prevalence of X-shaped magnetic fields likely associated with galactic winds \citep{krause_chang-es_2020}, further explored in \citet{stein_chang-es_2025}. Spectral ageing studies, especially in combination with low-frequency LOFAR data \citep{van_haarlem_lofar_2013}, allowed for detailed investigations of cosmic ray electron transport from the disc into the halo \citep{schmidt_chang-es_2019, stein_chang-es_2019, stein_chang-es_2019-1, miskolczi_chang-es_2019,heald_chang-es_2022, stein_chang-es_2023}. These results were supported by a scale height analysis by \citet{krause_chang-es_2018}, who showed that the escape of cosmic ray electrons is possible, indicating that the galaxies in the sample are non-calorimetric.\\
In this work, we build on the study of \citet{krause_chang-es_2018}, who measured scale heights of the thick radio disc in both $L$- and $C$-band. Instead of using these two bands, we make use of the new $S$-band data, which extends the existing CHANG-ES survey, as well as LOFAR observations of the galaxies at $\unit[144]{MHz}$. The addition of the $S$-band data enriches the CHANG-ES survey \citep{heesen_chang-es_2025}. This frequency band lies in the important transition region between Faraday-thin and Faraday-thick regimes. While we do not use the polarisation information in this study, we benefit from the $S$-band’s higher angular resolution and sensitivity. The LOFAR observations allow us to observe the same galaxy sample at much lower frequencies of $\unit[144]{MHz}$. These measurements taken as part of the LOFAR Two-metre Sky Survey \citep[LoTSS;][]{shimwell_lofar_2022}, roughly a factor of 20 lower in frequency than $S$-band allowing us to probe the cosmic ray electrons that have undergone significant spectral aging.\\
This paper is organised as follows:
Section 2 describes our sample selection, data reduction and the methodology. In Section 3 we present the results that are then discussed in Sec.~4. We conclude and give an outlook in Sec.~5. 

\section{Data and methodology}

\subsection{Data}

For this study, similar to \citet{heesen_chang-es_2025}, we use radio-continuum data observed with the Jansky Very Large Array (JVLA) in the $S$-band, i.e. in the frequency range of $2$--$4$\,GHz and data as part of the LOFAR Two-metre Sky Survey (LoTSS), i.e. in the frequency range of $\unit[144]{MHz}$. We used for the JVLA data the C-configuration, resulting in a nominal resolution of $7\arcsec$ using Brigg's robust weighting. The $S$-band data has a field of view of $15\arcmin$. The largest angular scale that could be imaged is $8\farcm 2$. Observations were done in the standard fashion with 2048 channels, each $\unit[100]{kHz}$ in bandwith. For the $S$-band data, we used the reduced data, which have also been used in \citet{heesen_chang-es_2025}.  
The LoTSS survey use the LOFAR High Band Antennas, using a frequency range of $120$--$168$\,MHz and relying exclusively on the Dutch array stations, yielding a nominal angular resolution of $6\arcsec$ \citep{shimwell_lofar_2022}. In its second data release (DR2), LoTSS covers about 27\% of the northern sky. For comparison with the JVLA data, we convolved the LoTSS maps to $7\arcsec$ resolution using {\sc imsmooth} within {\sc casa}.
We identified suitable mosaics in LoTSS-DR2 \citep{shimwell_lofar_2022} for the following 15 galaxies from the CHANG-ES sample: NGC~891, 2683, 2820, 3003, 3432, 3448, 4013, 4096, 4157, 4217, 4302, 4388, 4631, and 5907. Although NGC~3628 is not included in LoTSS-DR2, we identified a mosaic for this galaxy in LoTSS-DR3. The remaining CHANG-ES galaxies either lie at declinations too far south to be observed with LOFAR or lack mosaics of sufficient quality for our analysis.
In order to ensure the comparability between the JVLA $S$-band dataset and the data from LOFAR, we checked on inconsistencies in sensitivity and resolution between the two telescopes. For JVLA $S$-band in C-configuration, the largest angular scale is $490\arcsec$. There are four galaxies in the data sample which are more extended than the largest angular scale, these are NGC~891, 3628, 4631 and 5907. If there was missing flux, we would expect large scale height ratios for all of these galaxies. We found no trend indicating this. For LOFAR, the largest angular scale is not a limiting factor, as the shortest baselines are sufficiently small. To assess the sensitivity, we examined the extent of the measured profiles and found them to be comparable to those presented in \citet{heesen_chang-es_2025}. Furthermore, the measurement of scale heights is robust against moderate sensitivity variations, as they are derived from fitting procedures.

\subsection{Intensity profiles}

\begin{figure*}
    \centering
    \input{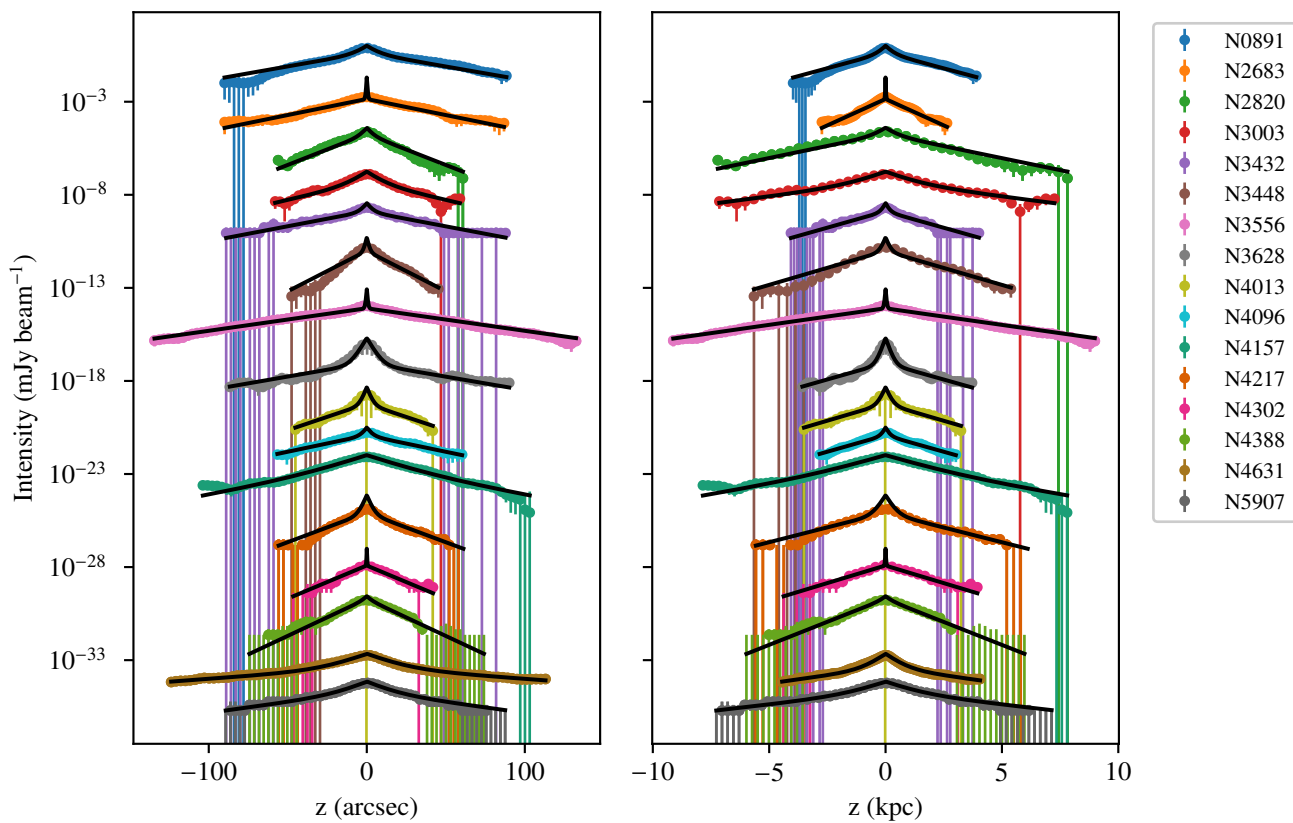}
    \caption{Showing vertical radio continuum intensity profiles at $\unit[144]{MHz}$ of the galaxies in our sample. We show the intensity profiles of the central strip for each galaxy, with exception of NGC~2820 and 4388 where the profile in the eastern strip is shown. The left panel shows the profiles in units of \unit[]{arcsec}, the right panel shows them in units of \unit[]{kpc} in projection of the assumed distances. Solid lines represent two-component exponential model profiles after deconvolution with the effective beam. The intensities were rescaled arbitrarily to separate the profiles for clearer visualization.}
    \label{fig:profiles}
\end{figure*}

\begin{figure}
    \centering
    \input{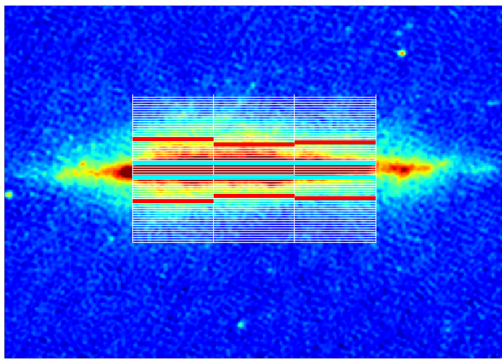}
    \caption{Location of the strips and boxes used for fitting vertical radio continuum intensity profiles overlaid on the LOFAR map of NGC~891. The red lines denote the measured scale heights of the thick disc for each strip, the blue lines denote the measured scale heights of the thin disc for each strip. We use three strips with a width in terms of the effective radius $r_e$ of $2/3\,r_{\rm e}$. For NGC~891, we calculate the average intensity in $60$ boxes per strip, each box with a height of $3\arcsec$, corresponding to approximately half of our angular resolution.}
    \label{fig:enter-label}
\end{figure}

We created vertical intensity profiles by averaging the emission within boxes arranged in strips perpendicular to the major axis of each galaxy. This approach follows the methodology described in \citet{krause_chang-es_2018}. For each galaxy, we defined three strips along the major axis: one centred on the galaxy and two placed symmetrically at larger galactocentric distances. In some cases, however, we could not derive scale heights in all strips due to limitations in data quality. For measuring the scale heights, we use the function {\sc boxmodels} of the program package {\sc nod3} \citep{muller_nod3_2017}. In order to create the intensity profiles required for this analysis, it is necessary to define the size and placement of the boxes used for averaging the emission. We set the total width of the three strips to correspond to the effective diameter (e-folding diameter) of the galaxy in the radio continuum, which we determined by placing a $10\arcsec$-wide strip along the major axis and fitting a Gaussian profile to the radio continuum emission to obtain the effective radius $r_{\rm e}$. Each strip was assigned a width of $2/3\,r_{\rm e}$. Within these strips, we averaged the intensities in boxes with widths equal to the strip width and heights of $3\arcsec$, corresponding to approximately half of our angular resolution. We fitted exponential functions to the vertical intensity profiles to determine the scale heights of the thin and thick discs. Since the angular resolution is limited, we need to account for the effective beam. The effective beam combines the clean beam (intrinsic resolution) and the projection of the inclined disc along the minor axis. The model profiles are thus the convolution of an exponential distribution with a Gaussian describing the effective beam, following the analytic approach by \citet{dumke_polarized_1995}. In Fig.~\ref{fig:profiles} we show vertical radio continuum profiles for the galaxies of our sample at LOFAR frequencies of $\unit[144]{MHz}$. Typically, we take the middle strip except NGC~2820 and 4388, where the middle strip had problems with proper fitting. These profiles reveal that the vertical emission cannot be described by a single exponential function. In the linear-log representation, a single exponential would appear as a straight line, yet our data clearly show changes in slope indicating more than one emission component. While some galaxies like NGC 3628 display sharp breaks between components, others such as NGC 4157 exhibit more gradual transitions. Thus, we can observe, both, the thin and thick discs and are able to separate them using double exponential fitting.
We choose to not apply a correction on the impact of the thermal emission as done for e.g. in \citet{stein_chang-es_2023}. \citet{stein_chang-es_2019} showed that non-thermal fractions for $C$-band lay in regions around $85\%$, for $L$-band around $95\%$. $S$-band might therefore lay somewhere between, and at LOFAR frequencies non-thermal fractions might be even higher. \citet{tabatabaei_radio_2017} finds similarly only small contributions of thermal emission \citep[see also][]{galvin_spectral_2018, dey_radio-only_2024}. Furthermore, we expect in the halo of the galaxies less thermal emission \citep{klein_radio_2018}. Even though, locally there might be lower non-thermal fractions \citep{irwin_chang-es_2024b}, global non-thermal fractions in the halo are usually high and the thermal emission provides only a minor contribution.We checked on the assumption that the thermal emission plays only a minor role in the thick disc by subtracting the thermal emission using maps from \citet{stein_chang-es_2023}. They show maps for five galaxies of our sample NGC~891, 3432, 4013, 4157 and 4631. The maps have a resolution of $20\arcsec$. For galaxies with scale heights larger than the resolution of the thermal maps, we find that the differences between scale heights measured on the total emission versus just the non-thermal emissions differs in average around $10\%$. This lays within the average error of the scale heights for these galaxies. For NGC~4013, the scale height of the thick disc is smaller than the resolution of the thermal maps. Measuring the scale height in that case leads to an unphysical high scale height. Therefore we see that in general the contribution of thermal emission in the thick disc is suppressed compared to the thin disc and the need for highly resolved \ion{H}{i} and $\mathrm{H\alpha}$ maps to correctly compute the thermal substraction.

\subsection{Integrated intensity profiles and scale height ratios}

In order to calculate the spectral index of the thick disc for the galaxy sample, we computed the integrated flux densities by integrating the intensities along the vertical direction. By using the effective radius of the galaxy, the radio continuum emission can be deprojected in order to obtain the mean deprojected intensity as it would appear in a face-on-galaxy.
Thus

\begin{equation}
I_i=w_i z_i/r_e,
\end{equation}
where $I_i$ is the deprojected mean intensity, $w_i$ is the amplitude in units of $\unit[]{mJy~beam^{-1}}$, $z_i$ is the scale height in units of $\unit[]{arcsec}$, and $r_{\rm e}$ is the effective radius in $\unit[]{arcsec}$. The index $i$ denotes the components of the radio emission, i.e. the thin and thick disc. Combining both $I=I_{\mathrm{thin}}+I_{\mathrm{thick}}$ yields the total mean intensity of the galaxy in the radio continuum. In what follows we only include results for the thick disc because the scale heights in the thin disc have too high relative errors in order to calculate ratios and spectral indices. In order to calculate the spectral index of the thick disc, we compute the deprojected mean intensity of the thick disc. 
The spectral index of the thick disc is then given by
\begin{equation}
    \alpha_2 = \frac{\log(I_{\nu_{1}}/I_{\nu_{2}})}{\log(\nu_1/\nu_2)} ,
\end{equation}
where $I_{\nu_{1}}$ and $I_{\nu_{2}}$ are the deprojected mean intensities at the frequencies $\nu_{1}$ and $\nu_{2}$, respectively – in our case $\nu_{1}=\unit[144]{MHz}$ from LOFAR and $\nu_{2}=\unit[3]{GHz}$ from the VLA $S$-band.
We measured the spectral index of the thick disc for each strip in every galaxy individually. The uncertainty of each spectral index was derived via error propagation of the measurement errors of the amplitude $w$ and the scale height $z$ in, both, the LOFAR and VLA $S$-band data. To obtain a global spectral index per galaxy, we combined the spectral indices measured in the individual strips using an inverse-variance weighting ($1/\sigma_i^2$). The weighted mean and the corresponding corrected standard deviation were calculated with the \texttt{DescrStatsW} routine from \texttt{statsmodels} \citep{seabold_statsmodels_2010}. \texttt{statsmodels} is a Python library which offers statistical tests, models, and data exploration, with \texttt{DescrStatsW} providing descriptive statistics such as weighted means and standard deviations for weighted datasets. An analogous procedure was applied to the ratio of the scale heights, determining the ratios for each strip and combining them into a global value for each galaxy. The weighted mean and its corrected standard deviation were again calculated with the \texttt{DescrStatsW} routine from \texttt{statsmodels}.

\section{Results}

\subsection{Scale height ratios}

\begin{figure}
    \centering
    \input{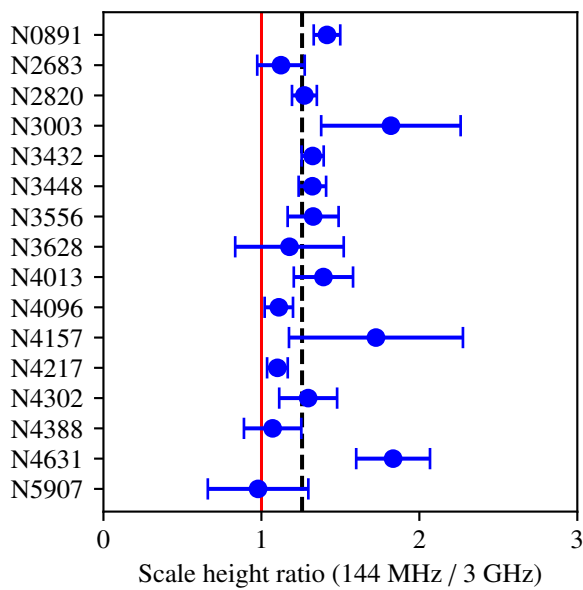}
    \caption{Ratio of scale heights of LOFAR observations at $\unit[144]{MHz}$ and VLA $S$-band observations with errors for all galaxies of the sample. The dotted black line indicates the mean scale height ratio of the galaxies in the sample.}
    \label{fig:shr_galaxies}
\end{figure}

First we investigate the scale height ratios of the thick radio disc between LOFAR observations at 144\,MHz and VLA $S$-band observations at 3\,GHz. The scale heights measured in the JVLA $S$-band for the set of CHANG-ES galaxies can be found in \citet{heesen_chang-es_2025}. We show the measured scale height ratios between JVLA $S$-band data and LoTSS data in Fig.~\ref{fig:shr_galaxies}. For the set of galaxies, we find a mean scale height ratio of $1.26\pm0.16$. With a ratio of $0.98\pm0.32$, NGC~5907 has the smallest ratio in the set of galaxies. Most galaxies have scale height ratios of below $1.5$. There are three galaxies with a scale height ratio bigger than $1.5$, the highest scale height ratio ($1.83\pm0.23$) can be found for NGC~4631.\\
In Fig.~\ref{fig:shr_prop}, we present relations between the scale height ratio and fundamental galaxy properties, such as total mass, star-forming radius, SFR surface density within $r_\star$, total mass surface density scaled to our distances, ratio of SFR-to-mass surface density and SFR. We tested for significant correlations between the different properties and the scale height ratio using Spearman’s rank correlation test ($p < 0.05$). Previous studies report correlations between these distinct galaxy properties and scale heights. \citet{krause_chang-es_2018} and \citet{galante_search_2024} found that there is a positive correlation between the radio-disc diameter and the scale height. \citet{heesen_chang-es_2025} also found a strong positive correlation between star forming radius and scale height. In \citet{heesen_chang-es_2025}, it also has been shown that for the thick disc there is a correlation between the SFR and the scale height, a negative correlation between the mass surface density and the scale height and a moderate positive correlation between SFR-to-surface density. However, in our analysis of scale height ratios, none of the investigated galaxy properties showed a statistically significant correlation with the scale height ratio. 

\begin{figure*}
    \centering
    \input{ratio_vs_properties_paper.pgf}
    \caption{Scale height ratio plotted against different galaxy properties. The panels show the scale height ratio versus total mass \citep{makarov_hyperleda_2014}, star forming radius\citep{wiegert_chang-es_2015}, SFR surface density within $r_\star$ \citep{heesen_chang-es_2025}, total mass surface density scaled to our distances \citep{irwin_continuum_2012}, ratio of SFR-to-mass surface density, SFR from $\mathrm{H\alpha}$ and mid-infrared \citep{vargas_chang-es_2019}. Data points represent the galaxies in our sample with their associated uncertainties.}
    \label{fig:shr_prop}
\end{figure*}

\subsection{Spectral index of the thick disc}

\begin{figure}
\centering
\input{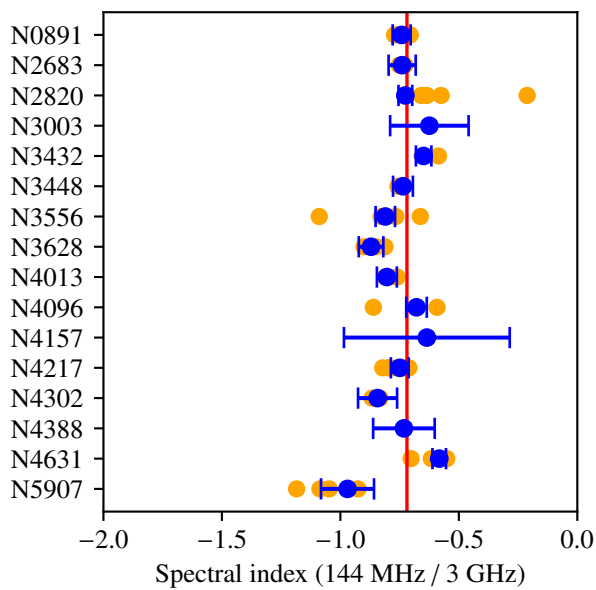}
\caption{Spectral index of the thick disc measured between LOFAR observations at $\unit[144]{MHz}$ and VLA $S$-band observations with errors for all galaxies of the sample. The red line represents the mean spectral index of the galaxy sample. The yellow points represent the measured spectral index in the different strips. The blue points (with error bars) denote the inverse-variance-weighted mean per strip and its standard error.}
\label{fig:spectral_thick}
\end{figure}

Here, we address the relations found between the spectral index measured in the thick disc, the fundamental galaxy quantities and the scale height ratio of the thick disc. By computing the integrated intensities, we are able to measure the spectral index between $\unit[144]{MHz}$ and $\unit[3]{GHz}$. In Fig.~\ref{fig:spectral_thick}, we present the spectral index of the thick disc measured in the galaxies. In our sample of CHANG-ES galaxies, we find an arithmetic mean spectral index of the thick disc of $-0.72\pm0.09$. The lowest spectral index is measured for NGC~5907 ($-0.97\pm0.11$). The highest spectral index is measured for NGC~4631 ($-0.58\pm0.03$). 
Overall, we find a spectral index still close to the injection spectral index, which is within the range expected when ionization or leakage losses play a role \citep{longair_high_2011}.\\

In Fig.~\ref{fig:alpha_prop}, we show the spectral index of the thick disc plotted against the same properties as for the scale heights. We find a significant correlation ($p$-value of $p<0.05$) for the total mass to spectral index. The correlation we find is strongly negative with $\rho_{\rm s}=-0.64$. Similar as we do, \citet{heesen_nearby_2022} also report a strong correlation between the total mass and the spectral index. We also find a significant correlation for the ratio of SFR-to-mass surface density and the spectral index. We find with $\rho_{\rm s}=0.51$ a strong positive correlation between both quantities. In addition, \citet{heesen_chang-es_2025} finds a moderate positive correlation for the SFR-to-mass surface density with scale height of the thick disc. There might also be a correlation between the SFR surface density within $r_\star$ and the spectral index ($p=0.07$, $\rho_{\rm s}=0.47$).
With $p$ being slightly higher than 0.05 the correlation might not be as significant as for the ratio of SFR-to-mass surface density. With $\rho_{\rm s}=0.47$, we see a moderate correlation strength between the SFR surface density within $r_\star$ and the spectral index. From the plotted data, one might speculate that the spectral index increases up to a value of about $\unit{10^{-2.5}}{\rm \,M_\sun\,yr^{-1}\,kpc^{-2}}$ and then remains approximately constant. \citet{tabatabaei_radio_2017} find a similar trend. The correlation shows that the cosmic ray electron (CRE) population in galaxies with higher SFR is younger and therefore the spectral index flat. Less star formation i.e. less supernova feedback leads to an aging population of CRE and therefore a steeper spectral index. However, more data are required to confirm this trend. 
For the star forming radius, there might be hints that there could be a weak correlation with the spectral index, but we are not able with our data to show a significant correlation ($\rho_s=-0.34$, $p=0.20$). We also find no correlation of the spectral index with the total mass-surface density ($\rho_s=-0.03$, $p=0.92$) or with the SFR ($\rho=-0.08$, $p=0.78$).

\begin{figure*}
    \centering
    \input{alpha_vs_properties_paper.pgf}
    \caption{Spectral index of the thick disc plotted against different galaxy properties. The panels show the spectral index versus total mass \citep{makarov_hyperleda_2014}, star forming radius\citep{wiegert_chang-es_2015}, SFR surface density within $r_\star$ \citep{heesen_chang-es_2025}, total mass surface density scaled to our distances \citep{irwin_continuum_2012}, ratio of SFR-to-mass surface density, SFR from $\mathrm{H\alpha}$ and mid-infrared \citep{vargas_chang-es_2019}. Data points represent the galaxies in our sample with their associated uncertainties.}
    \label{fig:alpha_prop}
\end{figure*}

\subsection{Comparing scale height ratio and spectral index of the thick disc}

\begin{figure}
    \centering
    \input{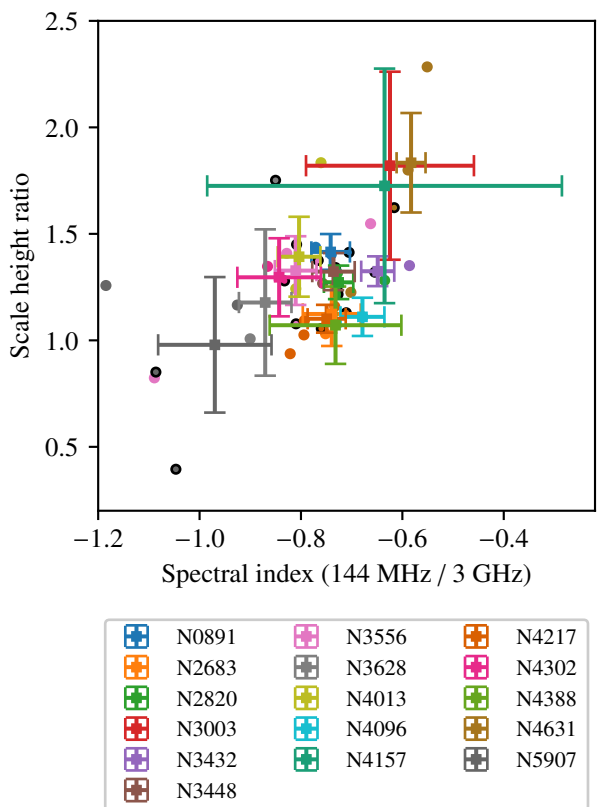}
    \caption{Scale height ratio plotted against the spectral index of the thick disc measured between LOFAR and VLA $S$-band data. Squares indicate the variance-weighted mean of all strips for each galaxy with the corresponding error bar, while dots show the individual strip measurements (errors omitted for clarity). The black circles indicate the central stip of the galaxy. If northern and southern halo are fittet individually, there are two central strips for the corresponding galaxy, and both are indicated in this plot.}
    \label{fig:alpha_shr}
\end{figure}

In this section, we combined both measurements of the scale height ratio and the spectral index of the thick disc. By combining both measurements, we expect to probe the influence of spectral aging on the extent of the thickness of the radio emitting halo at different frequencies. These combined observables provide constraints on the dominant CR transport processes. In Fig.~\ref{fig:alpha_shr}, we present the scale height ratio plotted against the spectral index of the thick disc. We show both the variance-weighted mean of all strips for each galaxy with the corresponding errors and the individual measurements of each strip.
We indicated the central strip of the galaxies with a black circle. For those galaxies where the northern and southern halo are fitted individually, there are two central strip fittings, both included in the plot.

We see a significant moderate correlation between both quantities. But opposite to our expectation, the correlation is positive and not negative, which we would expect if spectral aging was determining the scale height ratio. Measured only for the variance-weighted mean of all strips, we find a significant moderate positive correlation with $\rho_{\rm s}=0.49$ and $p=0.05$. Applying Spearman´s rank correlation test directly onto the measured values results in improved correlation coefficients of $\rho_{\rm s}=0.49$ and $p=6.5\times 10^{-4}$. 
For the central strips, however, we do not find any systematic trend; they are not consistently higher or lower than the remaining strips of a given galaxy.
The error bars on NGC~4157 might be high in the spectral index but still in combination with the data points of NGC~3003 and 4631 it seems plausible that there are galaxies which have a spectral index close to the injection spectral index and at the same time show a high scale height ratio. Although we find a moderate correlation, it might be useful to think the correlation in regimes of CR transport. There are galaxies in the sample which have high scale height ratios and flat spectral indices, most prominent NGC~3003, 4157, and 4631. \citet{stein_chang-es_2023} found similarly, as we do that for low frequencies the scale height in NGC~4631, that the one of NGC~4157 exceeds their high-frequency counterparts. \citet{stein_chang-es_2023} explain this behavior with advection-dominated galactic winds. As a consequence of winds playing a major role for these galaxies, the spectral index becomes flat. For NGC~4631 the gravitational interaction with NGC~4627 might be a reason for the strong winds. For NGC~3003 \citet{lu_edig-changes_2023} found high scale heights in $\mathrm{H\alpha}$ and \ion{H}{i} which fits in the picture of a extended thick disc with winds. \citet{stein_chang-es_2023} also reported enhanced low-frequency scale heights in NGC~891, again attributing them to the presence of a wind. In our analysis, NGC~891 likewise exhibits elevated values, though they do not quite reach the levels observed for NGC~3003, 4157, and 4631. On the other side of the parameter space, galaxies such as NGC~3628 and 5907 are located, characterized by steep spectral indices and comparatively low scale height ratios. We find for NGC~3628 and 5907 spectral indices fairly close to $\alpha=-1.1$. A integrated radio spectral index of $\alpha\approx-1.1$ is the theoretically expected integrated radio spectral index for a perfectly calorimetric galaxy \citep{lisenfeld_quantitative_1996}. Therefore we identify NGC~3628 and 5907 as being close to calorimetric halo. Scale height ratios close to unity suggest that even older CRE populations do not extend the thick disc further, implying that CREs are not escaping.
With an arithmetic mean value of $-0.72\pm0.09$, most galaxies lie between of the injection spectral index ($\alpha\approx-0.6$) \citep{lacki_diffuse_2013} and the spectral index of a calorimetric halo ($\alpha\approx-1.1$). This suggests that cosmic-ray electron transport along density gradients is plausible, which is also reflected in the scale height ratios of $1.26 \pm 0.16$.

\section{Discussion}

\subsection{Physical quantities of the individual galaxies}

Within our sample of galaxies, we find differences in the scale height ratios and spectral indices, which reflect variations in the underlying CRE transport. To understand the differences, we take a closer look on the individual galaxies in our sample. In the following, we will refer to the term spectral index of the thick disc simply as spectral index for brevity.\\
NGC~891 is one of the larger galaxies in the sample, with the third highest mass (Table~\ref{tab:galaxies}). \citet{stein_chang-es_2023} found that the galaxy shows extended low-frequency scale heights which indicates winds. Similarly, we also find an increased scale heights at low-frequencies for NGC~891, although the scale height is not as high as for example for NGC~3003, 4157, or 4631 which are also noted for their large scale heights in \citet{lu_edig-changes_2023, stein_chang-es_2023}. The spectral index of NGC~891 (Table~\ref{tab:galaxies}) is close to the mean spectral index in our sample of $-0.72\pm0.09$. Thus, unlike NGC~3003, 4157, or 4631, the spectral index of NGC~891 is not as close to the injection spectral index.\\
NGC~2683 is one of the galaxies in our sample that has an active galactic nucleus (AGN) \citep{heesen_nearby_2022}. Of all galaxies of our sample, NGC~2683 has the smallest star-forming radius, the highest total mass surface density, and the smallest SFR-to-mass surface density and SFR. The scale height ratio is slightly below the average of the sample. The spectral index is close to the average of the sample, so in between of aging and the calorimetric halo.\\
NGC~2820 represents an average case in our sample. Its scale height ratio is close to the average, the spectral index lies between the injection spectrum and the calorimetric limit, and the other galaxy properties also take values typical for our sample.\\
NGC~3003 has in our sample the second highest scale height ratio. \citet{lu_edig-changes_2023} showed that the galaxy shows even in $\mathrm{H\alpha}$ and \ion{H}{i} large scale heights, indicating an extended halo. NGC~2820 has a spectral index close to the injection spectral index and the lowest lowest SFR surface density within $r_\star$ of the galaxies in the sample.\\
NGC~3432 is the galaxy with the lowest total mass in our sample. It has a spectral index close to the injection spectral index and has scale height ratio slightly above the average of the sample but smaller than the scale height ratio of NGC~3003, 4157, and 4631.\\
NGC~3448 represents another average case in our sample. Its scale height ratio is close to the average, the spectral index lies between the injection spectrum and the calorimetric limit, and the other galaxy properties also take values typical for our sample.\\
NGC~3556 is one of the galaxies in our sample which might have an AGN \citep{satyapal_spitzer_2008}. It is one of the more massive galaxies in our sample and has the biggest SFR-to-mass surface density. The spectral index is steeper than the average spectral index. The scale height ratio is close to the average scale height. Interesting about this galaxy are the differences measured between the strips in the galaxy. There are two strips which show different behaviors. One strips shows low scale height ratios and a steep spectral index, the other strip shows a flat spectral index and high scale height ratios. This might be an indication that even inside of galaxies the transport mechanism spatially differ.\\
NGC~3628 is the second most massive galaxy in our sample and has the second-largest star-forming radius $r_\star$, both only surpassed by NGC~5907. It also exhibits the second-steepest spectral index and the second-lowest SFR surface density within $r_\star$, again after NGC~5907. The scale height ratio of NGC~3628 is smaller than the average scale height ratio of the sample. NGC~3628 also shows signs of gravitational interactions. \citet{li_chandra_2013} found X-ray emissions for this galaxy which suggest a large outflow-like halo, which appears surprising in connection with the steep spectral index. We find a higher scale height ratio for the central strip, so one might speculate about a nuclear outflow in the central strip.\\
NGC~4013 represents another average case in our sample. Its scale height ratio is close to the average, the spectral index lies between the injection spectrum and the calorimetric limit, and the other galaxy properties also take values typical for our sample.\\
NGC~4096 represents mostly another average case in our sample. The spectral index of NGC~4096 is slightly flatter than the average value. The scale height ratio is slightly smaller than the average value. The star-forming radius is comparatively small.\\
NGC~4157 has the third highest scale height ratio in combination with a flat spectral index, close to the injection spectral index. Similarly as do we, \citet{stein_chang-es_2023} found that the radio scale heights at low-frequencies exceed their high-frequency counterparts. The other galactic properties show typical values for our sample.\\
NGC~4217 represents another average case in our sample. Its scale height ratio is close to the average, the spectral index lies between the injection spectrum and the calorimetric limit, and the other galaxy properties also take values typical for our sample. For NGC~4217, \citet{heesen_chang-es_2024} found a $\unit[20]{kpc}$ radio bubble as third component besides the thin and thick disc. In our fit, we considered only the thin and thick disc.\\
NGC~4302 has a relatively steep spectral index, indicating spectral aging of CRE. The scale height ratio is close to the average value of the sample. NGC~4302 has the second lowest $\mathrm{\Sigma_{SFR}}$. The other properties show typical values for our sample. \citet{edler_victoria_2024} reported radio continuum tails in NGC~4302 and suggested that the galaxy might host an AGN. NGC~4302 interacts gravitationally with NGC~4298, \citet{zschaechner_investigating_2015} found a bridge connecting both galaxies.\\
NGC~4388 has a small scale height ratio close to 1. The spectral index is close to the average of the sample. NGC~4388 is a Seyfert-2 galaxy \citep{hummel_anomalous_1991}. We find no strong effect of the AGN on our results. In our sample NGC~4388 has the largest SFR surface density within $r_\star$ and the second largest SFR-to-mass surface density of the galaxies in the sample.\\
NGC~4631 has the highest scale height ratio and the flattest spectral index in our sample. NGC~4631 is influenced by tidal forces due to the interaction with its neighbor NGC~4627. The large scale heights at low-frequencies are also discussed in \citep{stein_chang-es_2023}. NGC~4631 has a relatively small ratio of SFR-to-mass surface density, the second highest SFR and the second lowest total mass surface density. \citet{mora-partiarroyo_chang-es_2019b} found in NGC~4631 large-scale magnetic field reversals.\\
NGC~5907 has a scale height ratio slightly lower than 1 and hence the smallest scale height ratio in the galaxy sample. With a spectral index close to $-1$, the galaxy has almost the theoretically expected spectral index for a perfectly calorimetric galaxy. NGC~5907 is the most massive galaxy of the sample, has the largest star-forming radius in the sample and the lowest SFR surface density within $r_\star$ of the sample.

\subsection{Galactic winds}

Galactic winds are extended outflows of cosmic rays, hot ionized plasma, cold neutral gas, and magnetic fields, driven by intense star formation and AGN \citep{thompson_theory_2024}. With galactic winds as the dominant transport mechanism, populations of cosmic-ray electrons are advected through the thick disc before synchrotron and inverse-Compton losses can significantly age the spectrum, resulting in a radio spectral index close to the injection value. In our sample of galaxies we find multiple galaxies which show a spectral index near the injection index ($\alpha \approx-0.6$), with galactic winds as plausible transport mechanism. Additionally for NGC~4631, the large-scale magnetic field reversals \citep{mora-partiarroyo_chang-es_2019b} might provide good conditions for reconnection heating \citep{wezgowiec_hot_2022}. In the case that large-scale bubbles rise into the halo, the adjacent magnetic field lines reconnect leading to increased efficiency in the transport of CRE \citep{mulcahy_resolved_2017}.
In our sample of galaxies, we also find systems with steep spectral indices, close to $\alpha \approx -1.1$. Such steep spectra are expected for calorimetric galaxies, in which cosmic-ray electrons are confined and radiative losses dominate \citep{lisenfeld_radio_2000}. For the calorimetric galaxies, we find additionally low scale height ratios. This supports the interpretation that cosmic rays are confined, reaching the thick disc but being prevented from escaping by radiative losses. In these calorimetric systems, a fountain-like behavior seems plausible. The cosmic-ray electrons reach certain height in the halo but eventually return, rather than escaping. This picture is reinforced by the fact that the galaxies we classify as calorimetric (e.g. NGC~3628 and 5907) are among the most massive in our sample.
The galaxies in our sample with galactic winds as dominant transport mechanism show relatively high scale height ratios. Although advection suppresses strong spectral ageing, the combination of radiative losses and vertical transport during the outflow produces the observed high scale height ratios. 
We therefore argue, that if cosmic ray transport happens, the dominant transport mechanism in our sample of galaxies is advection. Theoretically, ignoring magnetic field decline, the scale heights in galaxies with advective transport should follow the relation

\begin{equation}
    h \propto \nu^{-0.5} B^{-3/2},
\end{equation}
where $h$ is the scale height, $\nu$ is the frequency and $B$ is the magnetic field strength. Under the assumption of a constant magnetic field, the scale height fraction should, in the case of advective transport, only depend on the frequency ratio,

\begin{equation}
    \frac{h_{\nu_1}}{h_{\nu_2}}=\left(\frac{\nu_1}{\nu_2}\right)^{-0.5}.
\end{equation}
With a frequency ratio of roughly 20, we expect a scale height ratio of $\sim$$4.5$ in the case of advective transport. All galaxies in our sample show scale height ratios smaller than this value. 

A plausible explanation for smaller scale height ratios than expected from frequency ratios are dominant adiabatic losses. The synchrotron scale height is proportional to the product of the effective CRE lifetime and the advection speed.
When adiabatic losses dominate, where the timescales are energy independent, the effective CRE lifetime becomes nearly energy independent, resulting in similar scale heights at different frequencies and thus scale height ratios close to unity.
Therefore the scale height ratio decreases to values below the theoretical value, in our case $\sim4.5$. Diffusion does not appear to play a significant role in our sample. The vertical intensity distributions are well fitted by exponential profiles, while purely diffusive transport would yield Gaussian profiles \citep{heesen_advective_2016, stein_chang-es_2019}. This is also consistent with simulations, which show that diffusion is too weak to explain CR transport in extraplanar regions, where advection and streaming dominate \citep{hix_dynamically_2025}.
Recent studies based on multi-wavelength data emphasize the importance of pressure balance in driving galactic winds \citep{li_pressure_2024}. Especially the pressure equilibrium between the different phases, including the hot gas, cosmic rays, and magnetic fields, plays a crucial role in driving galactic winds. \citet{lu_edig-changes_2023} showed comparisons between scale heights and the different phases. They found that thermal and non-thermal electrons show almost the same spatial distributions. Thus, advection in galactic winds might be the dominant transport mechanism. Furthermore this indicates that the thermal gas, cosmic rays, and magnetic field may be close to energy equipartition. \citet{li_pressure_2024, li_edig-changes_2024} show for the galaxies NGC~3079 and 3556 the balance in pressure for the different phases.

When compared to \citet{heesen_nearby_2022}, we also find that spectral index depends on total mass with more massive galaxies having steeper spectra. Higher mass may imply a deeper gravitational potential, better confining CREs and leading to increased spectral aging. Furthermore, we find that galaxies with high SFR-to-mass surface densities have flatter spectra; these galaxies have also larger scale heights \citep{heesen_chang-es_2025}. A higher SFR-to-mass surface  ratio may indicate stronger feedback, driving advective winds which rapidly remove injected CREs, leading to a flattening of the spectrum. Both findings are in agreement with what would be expected for galactic winds. Similar to \citet{tabatabaei_radio_2017}, we find a correlation between SFR surface density within $r_\star$ and the spectral index. \citet{heesen_nearby_2022} did not find this relation. While \citet{heesen_nearby_2022} used the integrated spectral index, \citet{tabatabaei_radio_2017} uses the non-thermal spectral index and we use the spectral index of the thick disc. There might be absorption processes in the thin disc \citep{gajovic_spatially_2024}, obscuring this relation.

\section{Conclusions}

Cosmic rays and magnetic fields have a significant influence on the evolution and formation of galaxies. In order to understand the complex baryonic processes which regulate galaxy growth and evolution, the circumgalactic medium (CGM) is essential, as it hosts the key mechanisms of gas accretion, infall, phase transitions, and feedback that shape galaxy evolution. Radio continuum observations have been found valuable for probing the CGM \citep{heesen_nearby_2023}. Radio continuum observations of edge-on galaxies reveal two distinct discs. The thin disc, with a scale height of a few hundred parsec, consists mostly of stars, dust, and gas, while the thick disc contains ionized gas, cosmic rays, and magnetic fields \citep{ferriere_interstellar_2001, cox_three-phase_2005}. The latter, sometimes referred to as radio halo, may form the interface between the galaxy and the CGM where both outflows and accretion may be observed.

We investigated the vertical structure and spectral properties of the thick radio discs 
in edge-on galaxies from the CHANG-ES survey, combining low-frequency LOFAR LoTSS data at 144\,MHz 
with JVLA $S$-band measurements at 3\,GHz. We performed a two-component exponential fitting to the vertical intensity profiles in order to separate thin and thick disc. For the thick disc, we measured scale height ratios and spectral index between 144\,MHz and 3\,GHz.

The comparison of scale heights reveals a mean ratio of 
$1.26 \pm 0.16$, with most galaxies falling below a ratio of 1.5. The highest ratio is found in NGC\,4631 with $1.83\pm 0.23$, while NGC\,5907 shows a ratio close to unity with $0.98\pm 0.32$, suggesting little frequency-dependent 
thickening. The mean spectral index of the thick disc is 
$\langle \alpha_2 \rangle = -0.72 \pm 0.09$. This is fairly close to the injection spectral index possibly suggesting little spectral aging with fast electron escape.  We detect a flattening of the spectrum with increasing SFR-to-mass surface density 
ratio, and tentative evidence for a flatting with SFR surface density. In addition we found a strong negative correlation between total mass and spectral index, i.e. more massive galaxies show steeper spectra. The deeper gravitational potential may imply a better confinement of CREs and thereby increased spectral aging. 

One motivation for our work was to further investigate the radio continuum scale heights in the sample of \citet{heesen_chang-es_2025}. In their work it was shown that the scale height depends strongly on the SFR-to-mass surface density ratio. This can be either explained by a hydrostatic equilibrium for an atmosphere, or a galactic wind. These two cases can be distinguished using the frequency dependence of the scale heights. A high SFR-to-mass surface density ratio may indicate strong feedback and winds, rapidly removing injected CREs. Alternatively, there may be a galactic fountain with an effective mixing of halo gas.

Our work offers some new insights. We find a moderate positive correlation between scale height ratio and spectral index. Galaxies with flat spectra and high ratios, such as 
NGC\,3003, 4157, and 4631, are consistent with wind-driven cosmic-ray transport. Conversely, steep-spectrum systems, such as NGC\,5907 appear calorimetric, with 
scale heights insensitive to frequency, pointing to strong confinement of cosmic rays. While this suggests a connection between scale height ratios and galactic winds, none of the fundamental galaxy properties examined show a 
significant correlation with the scale height ratio. This is in contrast to previously reported 
trends for absolute scale heights. Yet, the flattening radio continuum spectrum with SFR-to-mass surface density ratio is indicative of winds, removing CREs before they age.

The measured scale height ratios are much lower than expected for either cosmic ray diffusion or advection if synchrotron and inverse Compton radiation losses dominate. For advection we would expect a ratio of $\sim$4.5 and for energy-independent diffusion a ratio of $\sim$2. Only for for a strongly energy-dependent diffusion, one would obtain the correct scale height ratio.
A plausible explanation for the small scale height ratios is that the CRE losses do not strongly depend on energy, such as is for escape via advection or adiabatic losses. Both are present in the case of accelerating galactic winds \citep[e.g.][]{heald_chang-es_2022}.

Therefore, we conclude that scale height ratios may be indeed a good proxy for cosmic ray transport and therefore galactic winds. However, it is yet unclear what drives galactic winds since there is no clear correlation between scale height ratio and any fundamental galaxy parameter. The spectral index may offer some clues, but it is confused with many effects particularly at low frequencies. We may see an influence of cosmic ray ionization losses playing a role \citep{gajovic_spatially_2024} and low-frequency free-free absorption \citep{gajovic_low-frequency_2025}. A flattening of spectra is also observed for global spectra in the gigahertz regime \citep{tabatabaei_radio_2017}, which was interpreted as the influence of star formation on the CRE population. It was not observed, however, in the global low frequency spectrum where no correlation between SFR surface density and spectral index was found \citep{heesen_nearby_2022}. The larger scale height of the radio continuum emission can be linked to a larger vertical scale height of the magnetic field or a larger scale height of CREs. The case of winds and CRE removal can be inferred via a flattening in the radio continuum spectrum as shown by \citet{tabatabaei_radio_2017} and tentatively this work.

Our findings demonstrate that scale height ratios and spectral indices are powerful tracers of cosmic-ray transport in galactic haloes. In the future, full radio SED modelling could help to distinguish between different loss mechanisms and to more quantitatively test cosmic-ray transport models.


\begin{acknowledgements}

We thank the anonymous referee for a concise and very helpful review.
DCS was supported by ErUM-IFT.  MB acknowledges funding by the Deutsche Forschungsgemeinschaft (DFG) under Germany's Excellence Strategy -- EXC 2121 ``Quantum Universe" --  390833306 and the DFG Research Group "Relativistic Jets". J.T.L. acknowledges the financial support from the China Manned Space Program with grant no. CMS-CSST-2025-A10 and CMS-CSST-2025-A04, and the National Science Foundation of China (NSFC) through the grants 12321003 and 12273111. TW acknowledges financial support from  the grant CEX2021-001131-S funded by MICIU/AEI/ 10.13039/501100011033, from the coordination of the participation in SKA-SPAIN, funded by the Ministry of Science, Innovation and Universities (MICIU). This research made use of following software packages and other resources: Aladin sky atlas developed at CDS, Strasbourg Observatory, France \citep{bonnarel_00a,boch_14a}; {\sc Astropy} \citep{astropy_13a,astropy_18a};  HyperLeda \citep[\href{http://leda.univ-lyon1.fr}{http://leda.univ-lyon1.fr};][]{makarov_14a}; NASA/IPAC Extragalactic Database (NED), which is operated by the Jet Propulsion Laboratory, California Institute of Technology, under contract with the National Aeronautics and Space Administration; SAOImage DS9 \citep{joye_03a}; and {\sc SciPy} \citep[\href{https://scipy.org}{https://scipy.org};][]{scipy_20a}.
\end{acknowledgements}

\bibliographystyle{aa}
\bibliography{lib}
\appendix

\section{Galaxy sample}
\setcounter{table}{0}
In Table~\ref{tab:galaxies} we present the scale height ratio, the spectral index and the properties of our sample galaxies.

\begin{table}[h]
\centering
\caption{Properties of the galaxies in the sample}
\label{tab:galaxies}
\begin{tabular}{lcccccccc}
\hline
 Galaxy   &  Scale height ratio  &  $\alpha_2$  &  \makecell{$\log_{10}(M_{\mathrm{tot}})$ \\ $[\mathrm{M}_\odot]$}  &  \makecell{$\log_{10}(r_\star)$ \\ $[{\rm kpc}]$}  &  \makecell{$\log_{10}(\Sigma_{\mathrm{SFR}})$ \\ $[\mathrm{M}_\odot\,{\rm yr}^{-1}\,{\rm kpc}^{-2}]$}  &  \makecell{$\log_{10}(\Sigma_{\mathrm{tot}})$ \\ $[\mathrm{M}_\odot\,{\rm kpc}^{-2}]$}  &  \makecell{$\log_{10}\!\left(\tfrac{\Sigma_{\mathrm{SFR}}}{\Sigma_{\mathrm{tot}}}\right)$ \\ $[{\rm yr}^{-1}]$}  &  \makecell{SFR \\ $[\mathrm{M}_\odot\,{\rm yr}^{-1}]$}  \\
\hline
 N0891    &     $1.42 \pm 0.08$      &   $-0.74 \pm 0.04$   &                           $11.12$                           &                        $1.10$                         &                                             $-2.42$                                             &                                      $8.39$                                      &                                                      $-10.81$                                                      &                      $1.88$                      \\
 N2683    &     $1.12 \pm 0.15$      &   $-0.74 \pm 0.06$   &                           $10.65$                           &                        $0.67$                        &                                             $-2.44$                                             &                                      $8.7$2                                      &                                                      $-11.16$                                                      &                      $0.25$                      \\
 N2820    &     $1.27 \pm 0.08$      &   $-0.73 \pm 0.03$   &                           $10.63$                           &                        $0.84$                        &                                             $-2.04$                                             &                                      $8.34$                                      &                                                      $-10.38$                                                      &                      $1.35$                      \\
 N3003    &     $1.82 \pm 0.44$      &   $-0.62 \pm 0.17$   &                           $10.67$                           &                        $1.14$                        &                                             $-2.59$                                             &                                      $7.77$                                      &                                                      $-10.35$                                                      &                      $1.56$                      \\
 N3432    &     $1.32 \pm 0.07$      &   $-0.65 \pm 0.03$   &                           $10.15$                           &                        $0.70$                         &                                             $-2.18$                                             &                                      $8.17$                                      &                                                      $-10.35$                                                      &                      $0.51$                      \\
 N3448    &     $1.32 \pm 0.09$      &   $-0.74 \pm 0.04$   &                           $10.33$                           &                        $0.79$                        &                                             $-1.84$                                             &                                      $8.42$                                      &                                                      $-10.25$                                                      &                      $1.78$                      \\
 N3556    &     $1.33 \pm 0.16$      &   $-0.81 \pm 0.04$   &                           $10.84$                           &                       $1.10$                        &                                             $-2.14$                                             &                                      $7.98$                                      &                                                      $-10.12$                                                      &                      $3.57$                      \\
 N3628    &     $1.18 \pm 0.34$      &   $-0.87 \pm 0.05$   &                           $11.13$                           &                        $1.09$                        &                                             $-2.52$                                             &                                      $8.32$                                      &                                                      $-10.85$                                                      &                      $1.41$                      \\
 N4013    &     $1.39 \pm 0.19$      &   $-0.80 \pm 0.04$   &                           $10.79$                           &                        $0.90$                         &                                             $-2.46$                                             &                                      $8.35$                                      &                                                      $-10.80$                                                       &                      $0.71$                      \\
 N4096    &     $1.11 \pm 0.09$      &   $-0.68 \pm 0.04$   &                           $10.46$                           &                        $0.77$                        &                                             $-2.19$                                             &                                      $8.34$                                      &                                                      $-10.53$                                                      &                      $0.71$                      \\
 N4157    &     $1.73 \pm 0.55$      &   $-0.64 \pm 0.35$   &                           $10.84$                           &                        $0.92$                        &                                             $-2.09$                                             &                                      $8.38$                                      &                                                      $-10.47$                                                      &                      $1.76$                      \\
 N4217    &     $1.10 \pm 0.07$      &   $-0.75 \pm 0.04$   &                           $10.99$                           &                        $1.07$                        &                                             $-2.36$                                             &                                      $8.28$                                      &                                                      $-10.64$                                                      &                      $1.89$                      \\
 N4302    &     $1.30 \pm 0.18$      &   $-0.84 \pm 0.08 $  &                           $10.85$                           &                        $1.03$                        &                                             $-2.60$                                              &                                      $8.25$                                      &                                                      $-10.85$                                                      &                      $0.92$                      \\
 N4388    &     $1.07 \pm 0.18$      &   $-0.73 \pm 0.13$   &                           $10.58$                           &                        $0.74 $                       &                                             $-1.60$                                              &                                      $8.57$                                      &                                                      $-10.17$                                                      &                      $2.42$                      \\
 N4631    &     $1.83 \pm 0.23$      &   $-0.58 \pm 0.03$   &                           $10.73$                           &                        $1.07$                        &                                             $-2.21$                                             &                                      $7.98$                                      &                                                      $-10.19$                                                      &                      $2.62$                      \\
 N5907    &     $0.98 \pm 0.32$      &   $-0.97 \pm 0.11$   &                           $11.33$                           &                        $1.25$                        &                                             $-2.66$                                             &                                      $8.13$                                      &                                                      $-10.79$                                                      &                      $2.21$                      \\
\hline
\end{tabular}
\end{table}

Table~\ref{tab:correlations} lists the Spearman correlation coefficients between the spectral index $\alpha_2$ and the scale height ratio with several galaxy properties. The table provides both the correlation coefficient $\rho_s$ and the corresponding $p$-value for each comparison. The data used to calculate these correlations are displayed in Figures~\ref{fig:shr_prop} and \ref{fig:alpha_prop}.

\begin{table}[h]
    \centering
    \caption{Spearman correlation coefficients between galaxy properties.}
    \label{tab:correlations}
    \begin{tabular}{lcccc}
    \hline
    
     Property & \multicolumn{2}{c}{$\alpha_2$} & \multicolumn{2}{c}{$\text{Scale height ratio}$} \\
    & $\rho_s$ & $p$ & $\rho_s$ & $p$ \\
    \hline
    $\log_{10}(M_{\mathrm{tot}})$ $[\mathrm{M}_\odot]$ 
        & $-0.64$ & $0.01$ & $-0.04$ & $0.87$ \\
    $\log_{10}(r_\star)$ $[{\rm kpc}]$ 
        & $-0.34$ & $0.20$ & $0.24$ & $0.36$ \\
    $\log_{10}(\Sigma_{\mathrm{SFR}})$ $[\mathrm{M}_\odot\,{\rm yr}^{-1}\,{\rm kpc}^{-2}]$ 
        & $0.47$ & $0.07$ & $0.02$ & $0.93$ \\
    $\log_{10}(\Sigma_{\mathrm{tot}})$ $[\mathrm{M}_\odot\,{\rm kpc}^{-2}]$ 
        & $-0.03$ & $0.92$ & $-0.27$ & $0.30$ \\
    $\log_{10}\!\left(\tfrac{\Sigma_{\mathrm{SFR}}}{\Sigma_{\mathrm{tot}}}\right)$ $[{\rm yr}^{-1}]$ 
        & $0.51$ & $0.04$ & $0.25$ & $0.35$ \\
    SFR  $[\mathrm{M}_\odot\,{\rm yr}^{-1}]$ 
        & $-0.08$ & $0.78$ & $0.06$ & $0.82$ \\
    \end{tabular}
\end{table}

\end{document}